\documentclass[sigconf]{acmart}

\acmSubmissionID{7935}

\usepackage{natbib}
\usepackage{algorithm}
\usepackage{multirow}
\usepackage[inline]{enumitem}
\usepackage{subfigure}
\usepackage{xcolor}
\usepackage{pifont}
\usepackage{acronym}
\usepackage{stfloats}

\copyrightyear{2023}
\acmYear{2023}
\setcopyright{acmlicensed}
\acmConference[WWW '23]{Proceedings of the ACM Web Conference 2023}{April 30-May 4, 2023}{Austin, TX, USA}
\acmBooktitle{Proceedings of the ACM Web Conference 2023 (WWW '23), April 30-May 4, 2023, Austin, TX, USA}
\acmPrice{15.00}
\acmDOI{10.1145/3543507.3583479}
\acmISBN{978-1-4503-9416-1/23/04}

\AtBeginDocument{%
  \providecommand\BibTeX{{%
    \normalfont B\kern-0.5em{\scshape i\kern-0.25em b}\kern-0.8em\TeX}}}

\acrodef{STEAM}{self-correcting sequential recommender}
\acrodef{RS}{recommender system}
\acrodef{SR}{sequential recommendation}
\acrodef{CV}{computer vision}
\acrodef{NLP}{natural language processing}
\acrodef{SSL}{self-supervised learning}
\acrodef{RL}{reinforcement learning}
\acrodef{MC}{Markov chain}
\acrodef{FM}{factorization machine}
\acrodef{MF}{matrix factorization}
\acrodef{RNN}{recurrent neural network}
\acrodef{LSTM}{long short-term memory}
\acrodef{GRU}{gated recurrent unit}
\acrodef{CNN}{convolutional neural network}
\acrodef{GNN}{graph neural network}
\acrodef{FFT}{fast Fourier transform}

\author{Yujie Lin}
\orcid{0000-0002-2146-0626}
\affiliation{%
\institution{Shandong University}
\city{Qingdao}
\country{China}
}
\email{yu.jie.lin@outlook.com}

\author{Chenyang Wang}
\orcid{0000-0001-6853-2116}
\affiliation{%
\institution{Shandong University}
\city{Qingdao}
\country{China}
}
\email{201900122032@mail.sdu.edu.cn}

\author{Zhumin Chen}
\orcid{0000-0003-4592-4074}
\affiliation{%
\institution{Shandong University}
\city{Qingdao}
\country{China}
}
\email{chenzhumin@sdu.edu.cn}

\author{Zhaochun Ren}
\orcid{0000-0002-9076-6565}
\affiliation{%
\institution{Shandong University}
\city{Qingdao}
\country{China}
}
\email{zhaochun.ren@sdu.edu.cn}

\author{Xin Xin}
\orcid{0000-0001-6116-9115}
\affiliation{%
\institution{Shandong University}
\city{Qingdao}
\country{China}
}
\email{xinxin@sdu.edu.cn}

\author{Qiang Yan}
\orcid{0000-0002-7328-2278}
\affiliation{%
\institution{WeChat, Tencent}
\city{Guangzhou}
\country{China}
}
\email{rolanyan@tencent.com}

\author{Maarten de Rijke}
\orcid{0000-0002-1086-0202}
\affiliation{%
\institution{University of Amsterdam}
\city{Amsterdam}
\country{The Netherlands}
}
\email{m.derijke@uva.nl}

\author{Xiuzhen Cheng}
\orcid{0000-0001-9998-2398}
\affiliation{%
\institution{Shandong University}
\city{Qingdao}
\country{China}
}
\email{xzcheng@sdu.edu.cn}

\author{Pengjie Ren}
\orcid{0000-0003-2964-6422}
\affiliation{%
\institution{Shandong University}
\city{Qingdao}
\country{China}
}
\email{renpengjie@sdu.edu.cn}
\authornote{Corresponding author.}

\renewcommand{\shortauthors}{Lin et al.}

\begin{document}

\title{A Self-Correcting Sequential Recommender}

\begin{abstract}
Sequential recommendations aim to capture users' preferences from their historical interactions so as to predict the next item that they will interact with.
Sequential recommendation methods usually assume that all items in a user's historical interactions reflect her/his preferences and transition patterns between items.
However, real-world interaction data is imperfect in that 
\begin{enumerate*}[label=(\roman*)]
\item users might erroneously click on items, i.e., so-called misclicks on irrelevant items, and 
\item users might miss items, i.e., unexposed relevant items due to inaccurate recommendations.
\end{enumerate*}

To tackle the two issues listed above, we propose \acs{STEAM}, a \textbf{S}elf-correc\textbf{T}ing s\textbf{E}quenti\textbf{A}l reco\textbf{M}mender.
\acs{STEAM} first corrects an input item sequence by adjusting the misclicked and/or missed items. It then uses the corrected item sequence to train a recommender and make the next item prediction. 
We design an item-wise corrector that can adaptively select one type of operation for each item in the sequence.
The operation types are `keep', `delete' and `insert.'
In order to train the item-wise corrector without requiring additional labeling, we design two self-supervised learning mechanisms:
\begin{enumerate*}[label=(\roman*)]
\item deletion correction (i.e., deleting randomly inserted items), and
\item insertion correction (i.e., predicting randomly deleted items). 
\end{enumerate*}
We integrate the corrector with the recommender by sharing the encoder and by training them jointly.
We conduct extensive experiments on three real-world datasets and the experimental results demonstrate that \acs{STEAM} outperforms state-of-the-art sequential recommendation baselines.
Our in-depth analyses confirm that \acs{STEAM} benefits from learning to correct the raw item sequences.
\end{abstract}



\maketitle

\acresetall


\section{Introduction}
\begin{figure}[t]
    \centering
    \includegraphics[width=\columnwidth]{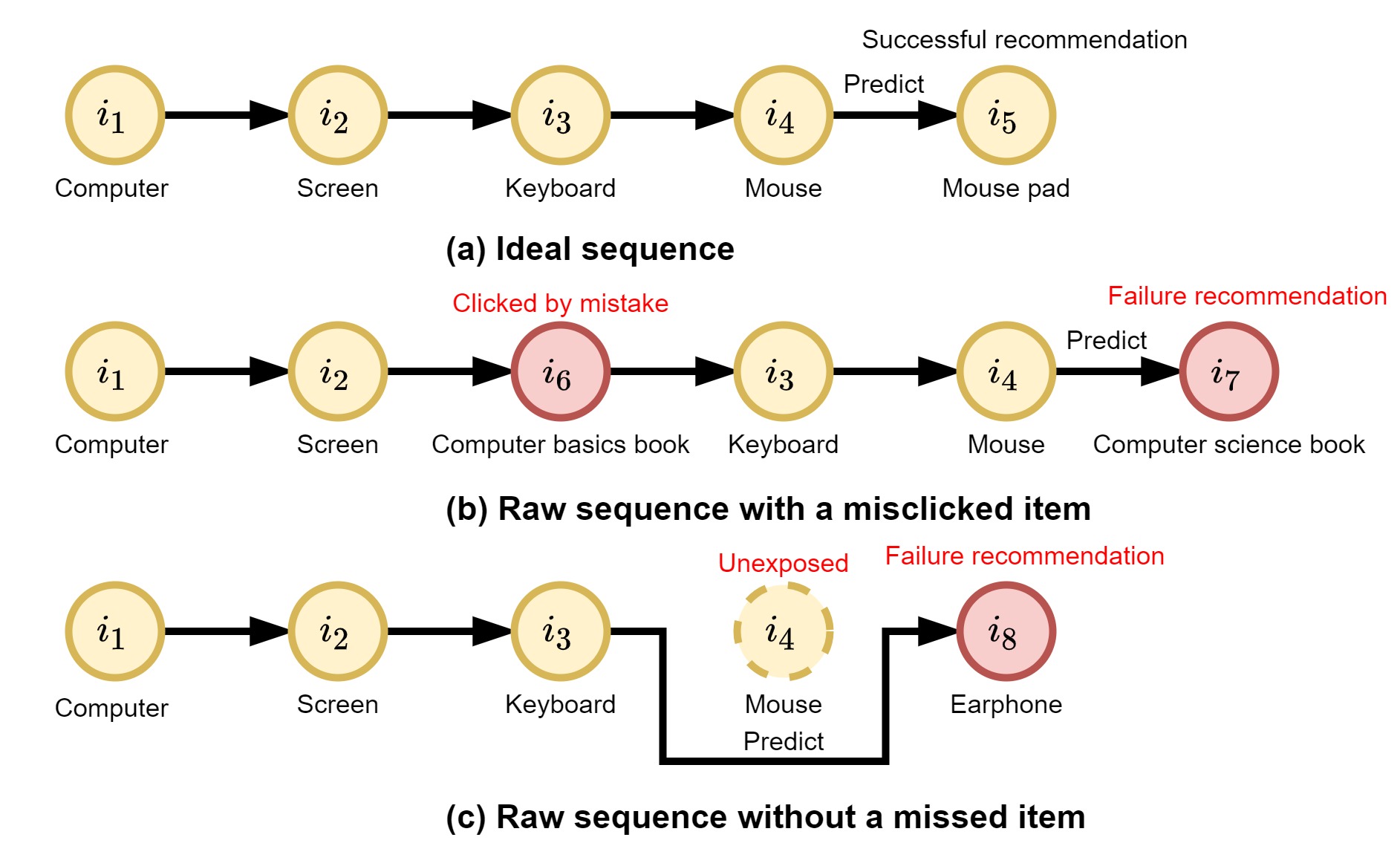}
    \caption{Examples for two kinds of imperfect item sequences. Sub-figure (a) is an ideal item sequence without any imperfection. Sub-figure (b) is an imperfect item sequence that contains a misclicked item (i.e., $i_6$). Sub-figure (c) is an imperfect sequence that lacks a missed item (i.e., $i_4$).}
    \Description{Examples for two kinds of imperfect item sequences. Sub-figure (a) is an ideal item sequence without any imperfection. Sub-figure (b) is an imperfect item sequence that contains a misclicked item (i.e., $i_6$). Sub-figure (c) is an imperfect sequence that lacks a missed item (i.e., $i_4$).}
    \label{figure_1_1}
\end{figure}

An important challenge of sequential recommendation is how to capture a user's preferences as accurately as possible by modeling the sequential dependencies of a user's historical interaction sequences~\cite{wang2019sequential,fang2020deep,wang2021survey}.
There is a considerable body of prior work towards this goal.
Early models are based on the \ac{MC} assumption that the next item only depends on its adjacent items to learn transition relationships \cite{rendle2010factorizing,garcin2013personalized,he2016fusing}.
Later, deep learning-based methods have been applied to sequential recommendation tasks for modeling more complex relations, such as \acp{RNN} \cite{wu2017recurrent,hidasi2018recurrent,ren2019repeatnet,ma2019pi}, \acp{CNN} \cite{tuan20173d,tang2018personalized,yuan2019simple}, memory networks \citep{chen2018sequential,huang2018improving,wang2019collaborative}, transformers \citep{kang2018self,sun2019bert4rec,wu2020sse}, and \acp{GNN} \citep{wu2019session,chang2021sequential,zhang2022dynamic}.
More recently, \ac{SSL} has been introduced to sequential recommendation for extracting robust item correlations by semi-automatically exploiting raw item sequences \citep{xia2021self,xie2022contrastive,yu2022self,ma2022improving}.

Most prior work ignores the fact that user-item interaction sequences may be imperfect, which means that they do not always accurately reflect user preferences and the transition patterns between items.
As illustrated in Fig.~\ref{figure_1_1}, there may be two kinds of imperfections in item sequences:
\begin{enumerate*}[label=(\roman*)]
\item The user may erroneously click on irrelevant items, so the item sequence may contain \emph{misclicked} items.
For example, in Fig.~\ref{figure_1_1}(b), the user mistakenly clicks $i_6$, which is a book about computer basics.
As a result, the recommendation model may recommend another book about computer science, i.e., $i_7$, which is actually not interesting for the user.
\item Some relevant items may not be exposed to the user, so the user may not be aware of them and thus will not click them.
As a result, the item sequence may lack some \emph{missed} items.
For instance, in Fig.~\ref{figure_1_1}(c), the user cannot interact with $i_4$, because $i_4$ is not exposed to the user.
Accordingly, the recommendation model may not further recommend $i_5$, a mouse pad, but may recommend $i_8$ which is an earphone.
\end{enumerate*}
For the first kind of imperfection, misclicked items can be considered as noise in item sequences.
Some studies have addressed capturing and eliminating noise from user-item interaction sequences \cite{qin2021world,zhou2022filter}.
These denoising sequential methods take the raw sequence as the input to train the model and predict the next item without explicitly modifying the given sequence.
Besides, they neglect the second kind of imperfection, which means they cannot recall missed and/or unexposed items.

Blindly learning a model on the raw data without considering its inherent imperfections may fail to capture a user’s true preferences, and harm the user experience and downgrade recommendation performance.
To this end, we enable a recommendation model to learn to correct the raw item sequence before making recommendations.
There are two main challenges to realize this correction.
First, the raw item sequence may mix different kinds of imperfections and contain multiple imperfections in multiple positions.
Hence, the first challenge is how to simultaneously apply different and multiple correction operations to the item sequence.
Second, it is difficult to identify the misclicked items and complement the missed items manually.
Therefore, the second challenge is how to train the corrector model without additional labeling.

We propose a novel sequential recommendation model, called \acfi{STEAM}, which first corrects the input item sequence using a corrector, and then uses the corrected item sequence to train a recommender and make the next item prediction.
Specifically, we propose an item-wise corrector that can adaptively apply `keep', `delete' and `insert' operations to the items in an input item sequence.
If the selected operation is `insert', the corrector will further use a reverse generator to generate the inserted sequence which will be inserted reversely before the item.
For misclicked items, the corrector can delete them, while for missed items, the corrector can insert them.
We design two self-supervised tasks to generate imperfect sequences and supervised signals automatically: 
\begin{enumerate*}[label=(\roman*)]
\item deletion correction, and 
\item insertion correction.
\end{enumerate*}
The former randomly inserts items to the raw item sequence and makes the model learn to delete them, while the latter randomly deletes items from the raw item sequence and recalls them afterwards.
We integrate the item-wise corrector and a recommender in \ac{STEAM} by sharing the encoder.
For the recommender, we use the raw item sequence and the corrected item sequence to train it by the masked item prediction task \cite{sun2019bert4rec}.
We use the joint loss from the corrector and the recommender to train the whole \ac{STEAM}.
We conduct experiments on three real-world datasets.
The results show that \ac{STEAM} significantly outperforms state-of-the-art sequential recommendation baselines.
We find that \ac{STEAM} benefits from learning to correct input sequences for better recommendation performance.
We also carry out experiments on simulated test sets by randomly inserting and deleting items, demonstrating that \ac{STEAM} is more robust than most baselines on more noisy data.

The main contributions of this work are as follows:
\begin{itemize}[leftmargin=*,nosep]
    \item We propose a \acf{STEAM} that is able to correct the raw item sequence before conducting recommendation.
    \item We design an item-wise corrector to correct the raw item sequence and two self-supervised learning mechanisms, deletion correction and insertion correction, to train the corrector.
    \item We conduct extensive experiments to demonstrate the state-of-the-art performance of \ac{STEAM}. To facilitate reproducibility, we release the code and data at \url{https://github.com/TempSDU/STEAM}.
\end{itemize}

\section{Related Work}

\subsection{Sequential recommendation}
Early work on sequential recommendation adopts \ac{MC} to capture the dynamic transition of user interactions.
\citet{rendle2010factorizing} combine first-order \acp{MC} and \ac{MF} to predict the subsequent user action.
\citet{he2016fusing} employ the high-order \acp{MC} to consider more preceding items and mine more complicated patterns.
With the development of deep learning, neural networks have been introduced to address sequential recommendation.
\citet{BalzsHidasi2016SessionbasedRW} adopt \acp{GRU} \cite{cho2014learning} to build a sequential recommendation model.
\citet{li2017neural} enhance the \ac{GRU}-based sequential recommendation model with an attention mechanism \cite{chaudhari2021attentive} to more accurately capture the user’s current preference.
\citet{tang2018personalized} propose a \ac{CNN}-based model to model sequential patterns in neighbors.
Later, more advanced neural networks have been applied.
\citet{chen2018sequential} introduce a memory mechanism \cite{weston2015memory} to design a memory-augmented neural network for leveraging users' historical records more effectively.
\citet{kang2018self} employ the unidirectional transformer \cite{vaswani2017attention} to capture long-range correlations between items.
\citet{sun2019bert4rec} further use a bidirectional transformer and the masked item prediction task for sequential recommendation.
\citet{wu2019session} utilize the \ac{GNN} \cite{welling2017semi} to model more complex item transition patterns in user sequences.

Recently, \acl{SSL} has demonstrated its effectiveness in extracting contextual features by constructing training signals from the raw data with dedicated tasks \cite{liu2021self}; it has been introduced to sequential recommendation as well.
\citet{zhou2020s3} propose four auxiliary self-supervised tasks to maximize the mutual information among attributes, items, and sequences.
\citet{xia2021self} propose to maximize the mutual information between sequence representations learned via hypergraph-based \acp{GNN}.
\citet{xie2022contrastive} use item crop, item mask, and item reorder as data augmentation approaches to construct self-supervision signals.
\citet{liu2021contrastive} propose data augmentation methods to construct self-supervised signals for better exploiting item correlations.
\citet{qiu2022contrastive} perform contrastive self-supervised learning based on dropout \cite{srivastava2014dropout}.

The studies listed above train models on the raw item sequences neglecting the fact that the raw sequences might be noisy due to users' casual click or inaccurate recommendations due to system exposure bias.

\subsection{Denoising recommendation}
Denoising recommendation aims to improve recommendation performance by alleviating the noisy data issue.
Prior work exploits additional user behavior and auxiliary item features to identify the noisy data, such as `skip' \cite{fox2005evaluating}, `dwell time' \cite{kim2014modeling}, `gaze' \cite{zhao2016gaze}, `like' \cite{bian2021denoising} and item side information \cite{lu2018between}.
These methods need extra feedback and manual labeling, which hinders their practical application.
Recently, another line of work has been dedicated to eliminating the effect of noisy data without introducing external signals.
\citet{wang2021denoising} observe that noisy data usually leads to large loss values in the early stage of training, and design two adaptive loss functions to down-weight noisy samples.
\citet{wang2021implicit} propose an iterative relabeling framework to identify the noise by exploiting the self-training principle.
\citet{wang2022learning} assume that predictions on noisy items vary across different recommendation models and propose an ensemble method to minimize the KL-divergence between the two models’ predictions.
\citet{SGDL22} argue that the models are prone to memorize easy and clean patterns at the early stage of training, so they collect memorized interactions at the early stage as guidance for the following training.

Most denoising recommendation methods focus on non-sequential recommendation.
There are a few studies targeting denoising for sequential recommendation.
\citet{qin2021world} design a denoising generator for next basket recommendation that is based on contrastive learning to determine whether an item in a historical basket is related to the target item.
\citet{tong2021pattern} mine sequential patterns as the prior knowledge to guide a contrastive policy learning model for denoising and recommendation.
Inspired by \acp{FFT} \cite{soliman1990continuous}, \citet{zhou2022filter} propose an all-MLP model with learnable filters for denoising sequential recommendation.

However, existing denoising recommendation methods do not correct the raw data explicitly.
Besides, they do not recall the missed items due to system exposure bias.

\section{Method}
\subsection{Overview}
\begin{figure*}[htbp]
    \centering
    \includegraphics[width=\linewidth]{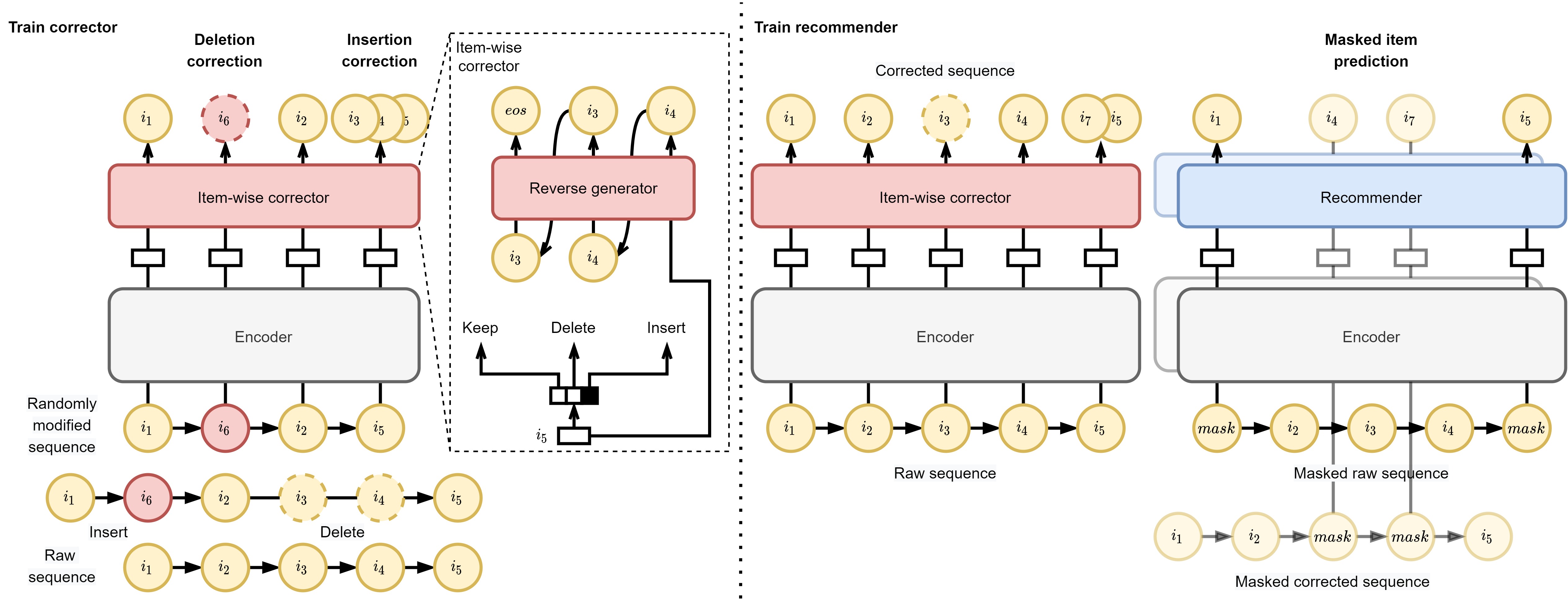}
    \caption{An overview of \ac{STEAM}. For training the corrector, the item-wise corrector is asked to perform deletion correction and insertion correction on items to recover the raw item sequence that has been randomly modified. The raw item sequence with its corrected version are both used to train the recommender using the masked item prediction task. Finally, \ac{STEAM} is optimized by the joint loss from the corrector and the recommender.}
    \Description{An overview of \ac{STEAM}. For training the corrector, the item-wise corrector is asked to perform deletion correction and insertion correction on items to recover the raw item sequence that has been randomly modified. The raw item sequence with its corrected version are both used to train the recommender using the masked item prediction task. Finally, \ac{STEAM} is optimized by the joint loss from the corrector and the recommender.}
    \label{figure_3_1}
\end{figure*}

We denote the item set as $\mathcal{I}$, and $|\mathcal{I}|$ is the number of items.
We denote an item sequence as $S=[i_1, \ldots, i_{|S|}]$, where $i_t \in \mathcal{I}$ is the interacted item at the $t$-th position of $S$ and $|S|$ is the sequence length.
We denote a subsequence $[i_j,\ldots,i_k]$ of $S$ as $S_{j:k}$.
Especially, we denote a raw item sequence as $S^r$.
For training \ac{STEAM} with deletion and insertion correction, we randomly modify $S^r$ and then ask \ac{STEAM} to recover it.
For each item in $S^r$, we keep it with probability $p_k$, insert one item before it with probability $p_i$, or delete it with probability $p_d$, where $p_k+p_i+p_d=1$.
Note that we can keep inserting more items with $p_i$, which are all sampled uniformly from $\mathcal{I}$$\setminus$$S$. 
For the last item in $S^r$, we would not delete it to avoid  confusion with the next item prediction.
The randomly modified sequence is denoted as $S^m$.
We have an operation sequence to mark the ground-truth correction operations on the items of $S^m$, denoted as $O=[o_1, \ldots, o_{|S^m|}]$, where $o \in $ \{`keep', `delete', `insert'\}.
Note that we do not consider `replace' as it can be achieved by the combination of `delete' and `insert.'
We denote all items in $S^m$ whose ground-truth correction operations are `insert' as $I^{ins}$.
For each item $i \in I^{ins}$, we denote the ground-truth inserted sequence as $S^{<i}=[i_1, \ldots, i_{|S^{<i}|-1}, [eos]]$, which should be inserted in reverse order before $i$, where $[eos] \in \mathcal{I}$ is a special token representing the ending.
Note that the order is $[i_{|S^{<i}|-1}, \ldots, i_1, i]$ after insertion.
We revise $S^r$ using the corrector to get the corrected sequence, which is denoted as $S^c$.
To train \ac{STEAM} with the masked item prediction, we randomly mask some items in $S^r$ or $S^c$ with a special token $[mask] \in \mathcal{I}$ with probability $p_m$, and then ask \ac{STEAM} to predict masked items.
The masked $S^r$($S^c$) and its masked items are denoted as $\widetilde{S}^r$($\widetilde{S}^c$) and $\widetilde{I}^r$($\widetilde{I}^c$).

The deletion and insertion correction tasks are to maximize $P(S^r|S^m)=P(O|S^m) \times \prod_{i \in I^{ins}}P(S^{<i}|S^m)$.
The masked item prediction task is to maximize $P(\widetilde{I}^r|\widetilde{S}^r)$ and $P(\widetilde{I}^c|\widetilde{S}^c)$.
The sequential recommendation task aims to predict the next item based on $P(i_{|S^r|+1}|S^r)$ or $P(i_{|S^c|+1}|S^c)$, which is equivalent to predict a masked item appending to the last position of $S^r$ or $S^c$.

An overview of \ac{STEAM} is shown in Fig.~\ref{figure_3_1}.
\ac{STEAM} has three main modules:
\begin{enumerate*}[label=(\roman*)]
\item a shared encoder, 
\item an item-wise corrector, and 
\item a recommender.
\end{enumerate*}
The encoder is used to encode the input sequence.
The item-wise corrector first predicts the correction operations on all items.
Then, the items whose correction operations are `delete' will be deleted.
For the items whose correction operations are `insert', the item-wise corrector uses the reverse generator to generate all inserted sequences.
The recommender aims to predict the masked items in item sequences.
To train the corrector, we take the randomly modified item sequence as input and ask the corrector to recover it.
To train the recommender, we first use the corrector to get the corrected item sequence.
Then, we randomly mask some items in the corrected item sequence or the raw item sequence, and use the recommender to predict them.
Finally, we use the joint loss to optimize \ac{STEAM}.
During testing, we append a $[mask]$ to the last position of the raw item sequence or the corrected item sequence, and use \ac{STEAM} to predict the next item.

Next, we provide the details of \ac{STEAM}.

\subsection{Encoder}
The target of the encoder is to encode the input item sequence to get the hidden representations of all positions of the item sequence.
The encoder is shared by the corrector and recommender, whose output will be the input of the two other modules.

Specifically, the encoder first maintains an item embedding matrix $\mathbf{E} \in \mathbb{R}^{e\times{|\mathcal{I}|}}$ to project the high-dimensional one-hot vector of an item to a low-dimensional dense vector.
For each item $i_t$ in a given sequence $S$, we follow Eq.~\ref{e_t} to project it:
\begin{equation}
\label{e_t}
\mathbf{e}_t = \mathbf{E}\mathbf{i}_t,
\end{equation}
where $\mathbf{i}_t \in \mathbb{R}^{|\mathcal{I}|}$ is the one-hot vector of $i_t$, $\mathbf{e}_t \in \mathbb{R}^e$ is the item embedding of $i_t$, and $e$ is the embedding size.
We inject the position information into the model by adding the position embeddings:
\begin{equation}
\label{h^0_t}
\mathbf{h}^0_t = \mathbf{e}_t+\mathbf{p}_t,
\end{equation}
where $\mathbf{p}_t \in \mathbb{R}^e$ is the learnable position embedding of the $t$-th position and $\mathbf{h}^0_t \in \mathbb{R}^e$ is the initial hidden representation of $i_t$.
Moreover, we follow \cite{kang2018self} to apply dropout to $\mathbf{h}^0_t$.
We further stack the initial hidden representations of all items in $S$ to get an initial hidden representation matrix $\mathbf{H}^0_e \in \mathbb{R}^{|S|\times{e}}$.

Then, the encoder employs a bidirectional transformer with $L_e$ layers to update $\mathbf{H}^0_e$, as shown in Eq.~\ref{H_e}:
\begin{equation}
\label{H_e}
\mathbf{H}^l_e = \mathrm{Trm_{bi}}(\mathbf{H}^{l-1}_e),
\end{equation}
where $\mathrm{Trm_{bi}}$ denotes a bidirectional transformer block; please refer to \citep{vaswani2017attention,sun2019bert4rec} for details. $\mathbf{H}^l_e \in \mathbb{R}^{|S|\times{e}}$ is the hidden representation matrix at the $l$-th layer.
$\mathbf{H}^{L_e}_e$ is the last hidden representation matrix, which will be the input of the item-wise corrector and the recommender.
For convenience, we ignore the superscript of $\mathbf{H}^{L_e}_e$ (i.e., $\mathbf{H}_e$) in the following modules.

\subsection{Item-wise corrector}
The item-wise corrector aims to execute correction operations at the item level. It selects one type of correction operation from `keep', `delete' and `insert' for each item in the raw item sequence.
If the selected correction operation is `insert' for an item, the item-wise corrector further uses a generator to generate the inserted sequence.

Given an item $i_t$ with its hidden representation $\mathbf{h}_t \in \mathbb{R}^e$ indexed from the input $\mathbf{H}_e$, we follow Eq.~\ref{p_o_t} to obtain the probability distribution $P(\hat{o}_t\mid S)$ for the corresponding correction operation $o_t$:
\begin{equation}
\label{p_o_t}
P(\hat{o}_t\mid S) = \mathrm{softmax}(\mathbf{W}\mathbf{h}_t),
\end{equation}
where $\hat{o}_t$ is the predicted version of $o_t$, $\mathbf{W} \in \mathbb{R}^{3\times{e}}$ is the projection matrix.
When testing, the operation with the maximum probability in $P(\hat{o}_t\mid S)$ will be applied to the item $i_t$.

Assuming that we have to insert items before the item $i_t$ of $S$ and its currently generated inserted sequence is $S^{<i_t}_{1:n-1}$, the reverse generator first follows the same way in the encoder to get the item embeddings of all items in $S^{<i_t}_{1:n-1}$.
Note that the item embedding matrix is shared between the encoder and the corrector.

Then, we stack the hidden representation $\mathbf{h}_t$ of $i_t$ obtained from $\mathbf{H}_e$ with all item embeddings $\{\mathbf{e}_1,\ldots,\mathbf{e}_{n-1}\}$ of $S^{<i_t}_{1:n-1}$:
\begin{equation}
\label{H^0_c}
\mathbf{H}^0_c = 
\begin{bmatrix} 
\mathbf{h}_t+\mathbf{p}_1 \\  
\mathbf{e}_1+\mathbf{p}_2 \\
\vdots \\
\mathbf{e}_{n-1}+\mathbf{p}_n
\end{bmatrix},
\end{equation}
where $\mathbf{H}^0_c \in \mathbb{R}^{n\times{e}}$ is the initial hidden representation matrix for the reverse generator. 
We also add the position embeddings here, which are shared between the encoder and the corrector too.
We apply dropout to $\mathbf{H}^0_c$ as we do in the encoder.

Next, the reverse generator uses a unidirectional transformer with $L_c$ layers to update $\mathbf{H}^0_c$ as Eq.~\ref{H_c}:
\begin{equation}
\label{H_c}
\mathbf{H}^l_c = \mathrm{Trm_{uni}}(\mathbf{H}^{l-1}_c),
\end{equation}
where $\mathrm{Trm_{uni}}$ represents a unidirectional transformer block; again, please see \citep{vaswani2017attention,kang2018self} for details.
$\mathbf{H}^l_c \in \mathbb{R}^{n\times{e}}$ is the hidden representation matrix at the $l$-th layer.
$\mathbf{H}^{L_c}_c $ is the last hidden representation matrix, and we denote it as $\mathbf{H}_c$ for short.

Finally, we calculate the probability distribution $P(\hat{i}_n\mid S^{<i_t}_{1:n-1},S)$ for the next inserted item $i_n$ by Eq.~\ref{p_i_n}:
\begin{equation}
\label{p_i_n}
P(\hat{i}_n\mid S^{<i_t}_{1:n-1},S) = \mathrm{softmax}(\mathbf{E}^{\top}\mathbf{h}_n),
\end{equation}
where $\mathbf{E}$ is the item embedding matrix, $\mathbf{h}_n \in \mathbb{R}^e$ is the hidden representation at the last position of $\mathbf{H}_c$.
In particulare, the first inserted item $i_1$ is generated based on $\mathbf{H}^0_c=[\mathbf{h}_t+\mathbf{p}_1]$, so we define $P(\hat{i}_1\mid S^{<i_t}_{1:0},S)=P(\hat{i}_1\mid S)$.

Since we have the complete ground-truth inserted sequence $S^{<i_t}$ when training, we can use the hidden representations of all positions of $\mathbf{H}_c$ to calculate $P(\hat{i}_{n+1}\mid S^{<i_t}_{1:n},S)$ for all $n$ at one time.
When testing, we will generate inserted items one by one with greedy search until generating $[eos]$ or achieving a maximum length.

\subsection{Recommender}
The recommender is to predict the masked items in $\widetilde{S}^r$ or $\widetilde{S}^c$, and to recommend the next item for $S^r$ or $S^c$.

Given the hidden representation matrix $\mathbf{H}_e$ of the input sequence $S$ from the encoder, we let $\mathbf{H}^0_r=\mathbf{H}_e$, where $\mathbf{H}^0_r \in \mathbb{R}^{|S|\times{e}}$ is the initial hidden representation matrix for the recommender.
First, the recommender utilizes a bidirectional transformer with $L_r$ layers to update $\mathbf{H}^0_r$, as shown in Eq.~\ref{H_r}:
\begin{equation}
\label{H_r}
\mathbf{H}^l_r = \mathrm{Trm_{bi}}(\mathbf{H}^{l-1}_r),
\end{equation}
where $\mathbf{H}^l_r \in \mathbb{R}^{|S|\times{e}}$ is the hidden representation matrix at the $l$-th layer.
$\mathbf{H}^{L_r}_r $ is the last hidden representation matrix, and we also denote it as $\mathbf{H}_r$ for short.

Then, assuming that we mask item $i_t$ in $S$ and we obtain its hidden representation $\mathbf{h}_t \in \mathbb{R}^e$ from $\mathbf{H}_r$, the recommender follows Eq.~\ref{p_i_t} to calculate the probability distribution $P(\hat{i}_t|S)$ for $i_t$:
\begin{equation}
\label{p_i_t}
P(\hat{i}_t\mid S) = \mathrm{softmax}(\mathbf{E}^{\top}\mathbf{h}_t),
\end{equation}
where $\mathbf{E}$ is the item embedding matrix, which is also used in the encoder and the corrector.
When training, the item $i_t \in \widetilde{I}^r$ $(\widetilde{I}^c)$ and the sequence $S$ is $\widetilde{S}^r$ $(\widetilde{S}^c)$.
When testing, the item $i_t$ is $i_{|S^r|+1}$ $(i_{|S^c|+1})$ and the sequence $S$ is $S^r(S^c)$, so we can get $P(i_{|S^r|+1}\mid S^r)$ ($P(i_{|S^c|+1}\mid S^c)$) for recommending the next item.

\subsection{Joint learning}
We train \ac{STEAM} with the deletion correction task, the insertion correction task, and the masked item prediction task.

For the correction tasks, we first randomly insert or delete items in a raw sequence $S^r$ to get a modified sequence $S^m$, then we ask \ac{STEAM} to delete the inserted items and insert the deleted items for $S^m$.
Through the deletion correction task and the insertion correction task, we can obtain self-supervised signals for correcting raw input sequences without additional manual labeling.
Specifically, our goal is to minimize the negative log-likelihood of $P(S^r\mid S^m)$, as shown in Eq.~\ref{L_1}:
\begin{equation}
\begin{split}
\label{L_1}
\mbox{}\hspace*{-10mm}
L_1 & \!=\! -\log P(S^r|S^m) \\
& \!=\! -\left(\log P(O|S^m)+\sum_{i \in I^{ins}}\log P(S^{<i}|S^m)\right) \\[-2pt]
& \!=\! -\left(\sum^{|S^m|}_{t=1}\log P(\hat{o}_t=o_t|S^m)\!+\!\!\sum_{i \in I^{ins}}\!\sum^{|S^{<i}|}_{n=1}\log P(\hat{i}_n=i_n|S^{<i}_{1:n-1},S^m)\!\right),
\hspace*{-10mm}\mbox{}
\end{split}
\end{equation}
where $L_1$ is the loss for the corrector.

In the masked item prediction task, we first employ \ac{STEAM} to correct $S^r$ to get the corrected item sequence $S^c$, then randomly mask items in $S^r$ and $S^c$, and use \ac{STEAM} to predict the masked items.
We also minimize the negative log-likelihood of $P(\widetilde{I}^r\mid \widetilde{S}^r)$ and $P(\widetilde{I}^c\mid \widetilde{S}^c)$, i.e., Eq.~\ref{L_2}:
\begin{equation}
\begin{split}
\label{L_2}
L_2 & = -\left(\log P(\widetilde{I}^r\mid \widetilde{S}^r)+\log P(\widetilde{I}^c\mid \widetilde{S}^c)\right) \\
& = -\left(\sum_{i \in \widetilde{I}^r}\log P(\hat{i}=i\mid \widetilde{S}^r)+\sum_{i \in \widetilde{I}^c}\log P(\hat{i}=i\mid \widetilde{S}^c)\right),
\end{split}
\end{equation}
where $L_2$ is the loss for the recommender.

Finally, we use the joint loss $L$ shown in Eq.~\ref{L} to optimize the parameters of \ac{STEAM}:
\begin{equation}
\label{L}
L = L_1+L_2.
\end{equation}
Here, $L$ is minimized by the standard backpropagation algorithm.


\section{Experimental Setup}
\subsection{Research questions}
We seek to answer the following research questions:
\begin{enumerate*}[label=\textbf{(RQ\arabic*)},leftmargin=*,nosep]
    \item How does \ac{STEAM} perform compared with the state-of-the-art sequential recommendation methods?
    \item What benefits can the recommender in \ac{STEAM} obtain from the corrector in \ac{STEAM}?
    \item How does \ac{STEAM} perform with different levels of noise?
\end{enumerate*}

\subsection{Datasets}
\label{section:datasets}
We evaluate \ac{STEAM} on three datasets with varying domains.
\textbf{Beauty} and \textbf{Sports} belong to a series of product review datasets crawled from Amazon.com by \citet{mcauley2015image}.
We select Beauty and Sports subcategories in this work following \citet{zhou2022filter}.
\textbf{Yelp} is a dataset for business recommendation released by Yelp.com.
As it is very large, we only use the transaction records between ``2019-01-01'' and ``2019-12-31.''

We follow common practices \cite{kang2018self,zhou2022filter} to preprocess all datasets.
We remove users and items whose interactions are less than $5$.
We sort each user's interactions by time to construct an item sequence.
For each item sequence, we use the last item for testing, the second last item for validation, and the remaining items for training.
We pair the ground-truth item for testing or validation with $99$ randomly sampled negative items that the user has not interacted with.

For each dataset, we also construct a \emph{simulated} test set, which is different from the real test set by randomly modifying test sequences to introduce more imperfections.
For each item in a test sequence excluding the ground-truth item, we will keep it with probability $0.8$, insert one item before it (simulating misclicks on irrelevant items) or delete it (simulating unexposed relevant items due to inaccurate recommendations) with  probability $0.1$.
We limit the count of continuous insertion operations less than $5$. 
The inserted item is sampled uniformly from the item set.
The statistics of the processed datasets are summarized in Table~\ref{dataset}.

\begin{table}[h]
\caption{Statistics of the datasets after preprocessing.}
\Description{Statistics of the datasets after preprocessing.}
\label{dataset}
\small
\begin{tabular}{l ccccc}
\toprule
Dataset & \#Users & \#Items & \#Actions & Avg. length & Sparsity \\ 
\midrule
Beauty  & 22,362  & 12,101  & 194,682   & \phantom{1}8.7         & 99.93\%  \\
Sports  & 35,597  & 18,357  & 294,483   & \phantom{1}8.3         & 99.95\%  \\
Yelp    & 22,844  & 16,552  & 236,999   & 10.4        & 99.94\%  \\
\bottomrule
\end{tabular}
\end{table}

\subsection{Baselines}
\label{section:baselines}
We compare \ac{STEAM} with the following representative baselines, which can be grouped into
\begin{enumerate*}[label=(\roman*)]
\item vanilla sequential recommendation models, 
\item SSL-based sequential recommendation models, and
\item denoising sequential recommendation models.
\end{enumerate*}
For each group, we only consider the recent state-of-the-art methods.
\begin{itemize}[leftmargin=*,nosep]
\item \textbf{Vanilla sequential recommendation models:}
    \begin{itemize}
    \item \textbf{GRU4Rec} \cite{BalzsHidasi2016SessionbasedRW} employs a \ac{GRU} to model sequential patterns between items for sequential recommendation.
    \item \textbf{SASRec} \cite{kang2018self} uses a unidirectional transformer to model item sequences for predicting next items.
    \item \textbf{BERT4Rec} \cite{sun2019bert4rec} adopts a bidirectional transformer trained on the masked item prediction task.
    \item \textbf{SRGNN} \cite{wu2019session} models item sequences using a \ac{GNN} with an attention network.
    \end{itemize}
\item \textbf{SSL-based sequential recommendation models:}
    \begin{itemize}
    \item \textbf{CL4SRec} \cite{xie2022contrastive} uses three self-supervised tasks based on item crop, item mask, and item reorder respectively to train a transfor-mer-based sequential recommendation model.
    \item \textbf{DuoRec} \cite{qiu2022contrastive} is the state-of-the-art SSL-based sequential method that employs a model-level augmentation approach based on dropout and a novel sampling strategy to construct contrastive self-supervised signals.
    \end{itemize}
\item \textbf{Denoising sequential recommendation models:}
    \begin{itemize}
    \item \textbf{FMLP-Rec} \cite{zhou2022filter} integrates \ac{FFT} with an all-MLP architecture for denoising in sequential recommendation. There are few research on denoising for sequential recommendation work. We select FMLP-Rec as it is the latest state-of-the-art one.
    \end{itemize}
\end{itemize}
We also report ``Recommender'', which is a variant of the recommender in \ac{STEAM} that is trained by the masked item prediction task without joint training with the corrector.
For \ac{STEAM}, we report its performance on corrected item sequences by default.
\balance

\subsection{Metrics and implementation}
We adopt two widely used evaluation metrics to evaluate the performances of all sequential recommendation methods: HR@$k$ (hit ratio) and MRR@$k$ (mean reciprocal rank) \cite{fang2020deep}, where $k \in \{5,10\}$.

For all baselines and \ac{STEAM}, we initialize the trainable parameters randomly with Xavier method \cite{glorot2010understanding}.
We optimize all methods with the Adam optimizer \cite{kingma2015adam} for $300$ epochs, with a learning rate of $0.001$ and a batch size of $256$.
We also apply gradient clipping \cite{pascanu2013difficulty} with range $[-5,5]$ during training.
We set the maximum raw item sequence length to $50$, the maximum item sequence length after correction to $60$, and the maximum number of continuous inserted items to $5$.

For all baselines, we follow the instructions from their original papers to set the hyper-parameters.
For the hyper-parameters of \ac{STEAM}, we set the embedding size $e$ to $64$, the number of heads in transformer to $1$, and the number of layers $L_e$, $L_c$ and $L_r$ to $1$.
We set the dropout rate to $0.5$.
During training, the keep probability $p_k$, insertion probability $p_i$, deletion probability $p_d$, and mask probability $p_m$ are set to $0.4$, $0.1$, $0.5$, and $0.5$, respectively.

\section{Experimental Results}
\begin{table*}[htbp]
\caption{Performance comparison of different methods on the real test sets. The best performance and the second best performance are denoted in bold and underlined fonts respectively. $*$ indicates that the performance gain of \ac{STEAM} against the best baseline is statistically significant based on a two-sided paired t-test with $p<0.05$.}
\Description{Performance comparison of different methods on the real test sets. The best performance and the second best performance are denoted in bold and underlined fonts respectively. $*$ indicates that the performance gain of \ac{STEAM} against the best baseline is statistically significant based on a two-sided paired t-test with $p<0.05$.}
\label{table_5_1}
\small
\setlength{\tabcolsep}{1mm}
\begin{tabular}{l cccc cccc cccc}
\toprule
                       & \multicolumn{4}{c}{Real Beauty}                                                         & \multicolumn{4}{c}{Real Sports}                                                                               & \multicolumn{4}{c}{Real Yelp}                                                                                 \\
\cmidrule(r){2-5} \cmidrule(r){6-9} \cmidrule(r){10-13}                          
Model                  & HR@5                      & HR@10                     & MRR@5          & MRR@10         & HR@5                      & HR@10                     & MRR@5                     & MRR@10                    & HR@5                      & HR@10                     & MRR@5                     & MRR@10                    \\
\midrule
GRU4Rec                & 32.95                     & 42.59                     & 21.63          & 22.90          & 30.58                     & 42.85                     & 18.35                     & 19.97                     & 55.40                     & 76.57                     & 32.23                     & 35.05                     \\
SASRec                 & 36.58                     & 45.57                     & 25.43          & 26.62          & 34.51                     & 46.20                     & 21.91                     & 23.46                     & 58.24                     & 77.96                     & 35.07                     & 37.72                     \\
BERT4Rec               & 36.67                     & 47.28                     & 23.38          & 24.79          & 35.16                     & 47.91                     & 21.54                     & 23.24                     & 61.18                     & 79.72                     & 37.64                     & 40.13                     \\
SRGNN                  & 37.33                     & 47.65                     & 25.15          & 26.52          & 35.92                     & 48.32                     & 22.44                     & 24.08                     & 59.86                     & 78.96                     & 36.74                     & 39.30                     \\
\midrule
CL4SRec                & 39.29                     & 48.75                     & 27.59          & 28.84          & 37.91                     & 49.83                     & 24.53                     & 26.11                     & 62.15                     & 80.16                     & 39.29                     & 41.70                     \\
DuoRec                 & \underline{40.95}               & \underline{50.78}               & \textbf{28.84} & \textbf{30.15} & \underline{39.80}               & \underline{51.93}               & \underline{25.97}               & \underline{27.58}               & \underline{64.01}               & \underline{82.63}               & \underline{40.85}               & \underline{43.34}               \\
\midrule
FMLP-Rec               & 39.69                     & 48.72                     & 28.01          & 29.20          & 37.67                     & 49.32                     & 24.66                     & 26.21                     & 61.85                     & 80.76                     & 38.38                     & 40.92                     \\
\midrule
Recommender            & 35.73                     & 46.47                     & 22.84          & 24.27          & 35.02                     & 47.78                     & 21.34                     & 23.03                     & 61.41                     & 80.57                     & 37.67                     & 40.22                     \\
\ac{STEAM}             & \textbf{42.57}\rlap{$^*$} & \textbf{52.89}\rlap{$^*$} & \underline{28.75}    & \underline{30.14}    & \textbf{42.14}\rlap{$^*$} & \textbf{55.16}\rlap{$^*$} & \textbf{26.87}\rlap{$^*$} & \textbf{28.61}\rlap{$^*$} & \textbf{67.22}\rlap{$^*$} & \textbf{84.49}\rlap{$^*$} & \textbf{43.45}\rlap{$^*$} & \textbf{45.77}\rlap{$^*$} \\
\bottomrule
\end{tabular}
\end{table*}

\begin{table*}[htbp]
\caption{Performance analysis of \ac{STEAM} on different groups of the real test sets. Overall-R (Overall-C) is the performance on all raw (corrected) test item sequences. Changed-R (Changed-C) is the performance on the raw (corrected) test item sequences of the changed sequence group. Unchanged is the performance on the test item sequences of the unchanged sequence group.}
\Description{Performance analysis of \ac{STEAM} on different groups of the real test sets. Overall-R (Overall-C) is the performance on all raw (corrected) test item sequences. Changed-R (Changed-C) is the performance on the raw (corrected) test item sequences of the changed sequence group. Unchanged is the performance on the test item sequences of the unchanged sequence group.}
\label{table_5_2}
\small
\setlength{\tabcolsep}{1mm}
\begin{tabular}{l cccc cccc cccc}
\toprule
                            & \multicolumn{4}{c}{Real Beauty} & \multicolumn{4}{c}{Real Sports} & \multicolumn{4}{c}{Real Yelp}  \\
\cmidrule(r){2-5} \cmidrule(r){6-9} \cmidrule(r){10-13}
\ac{STEAM}                  & HR@5   & HR@10 & MRR@5 & MRR@10 & HR@5   & HR@10 & MRR@5 & MRR@10 & HR@5  & HR@10 & MRR@5 & MRR@10 \\
\midrule
Overall-R                   & 42.21  & 52.75 & 28.27 & 29.68  & 42.03  & 55.04 & 26.75 & 28.48  & 67.19 & 84.49 & 43.42 & 45.75  \\
Overall-C                   & 42.57  & 52.89 & 28.75 & 30.14  & 42.14  & 55.16 & 26.87 & 28.61  & 67.22 & 84.49 & 43.45 & 45.77  \\
\midrule
Changed-R                   & 41.35  & 51.59 & 27.04 & 28.40  & 35.04  & 47.64 & 21.56 & 23.23  & 56.05 & 74.19 & 34.36 & 36.80  \\
Changed-C                   & 42.56  & 52.06 & 28.66 & 29.94  & 35.54  & 48.12 & 22.08 & 23.76  & 57.46 & 74.40 & 35.45 & 37.73  \\
\midrule
Unchanged                   & 42.58  & 53.25 & 28.79 & 30.22  & 44.21  & 57.36 & 28.37 & 30.12  & 67.44 & 84.71 & 43.62 & 45.95  \\ 
\bottomrule
\end{tabular}
\end{table*}

\begin{table}[htbp]
\caption{Statistics of correction operations by \ac{STEAM} on the real test sets. \#Changed is the percentage of the changed test item sequences after correction. \#Keep, \#Delete and \#Insert are the percentages of different types of correction operations during correction.}
\Description{Statistics of correction operations by \ac{STEAM} on the real test sets. \#Changed is the percentage of the changed test item sequences after correction. \#Keep, \#Delete and \#Insert are the percentages of different types of correction operations during correction.}
\label{table_5_3}
\small
\begin{tabular}{l c ccc}
\toprule
Dataset     & \#Changed       & \#Keep & \#Delete & \#Insert \\
\midrule
Real Beauty & 29.91           & 88.60  & 4.03     & 7.37     \\
Real Sports & 23.82           & 95.72  & 4.21     & 0.07     \\
Real Yelp   & \phantom{0}2.17 & 99.63  & 0.15     & 0.22     \\    
\bottomrule
\end{tabular}
\end{table}

\begin{table*}[t]
\caption{Performance comparison of different methods on the simulated test sets.}
\Description{Performance comparison of different methods on the simulated test sets.}
\label{table_5_4}
\small
\setlength{\tabcolsep}{1mm}
\begin{tabular}{l cccc cccc cccc}
\toprule
                       & \multicolumn{4}{c}{Simulated Beauty}                                                    & \multicolumn{4}{c}{Simulated Sports}                                                                          & \multicolumn{4}{c}{Simulated Yelp}                                                              
            \\
\cmidrule(r){2-5} \cmidrule(r){6-9} \cmidrule(r){10-13}
Model                  & HR@5                      & HR@10                     & MRR@5          & MRR@10         & HR@5                      & HR@10                     & MRR@5                     & MRR@10                    & HR@5                      & HR@10                     & MRR@5                     & MRR@10                    \\
\midrule
GRU4Rec                & 32.22                     & 42.13                     & 21.28          & 22.59          & 29.96                     & 42.26                     & 17.99                     & 19.61                     & 54.64                     & 75.87                     & 31.66                     & 34.49                     \\
SASRec                 & 35.97                     & 45.26                     & 24.97          & 26.20          & 33.63                     & 45.23                     & 21.47                     & 23.01                     & 57.71                     & 77.12                     & 34.64                     & 37.23                     \\
BERT4Rec               & 35.83                     & 46.79                     & 22.79          & 24.25          & 34.10                     & 46.49                     & 20.62                     & 22.26                     & 59.46                     & 78.07                     & 36.36                     & 38.85                     \\
SRGNN                  & 36.64                     & 46.81                     & 24.50          & 25.85          & 35.39                     & 47.55                     & 22.00                     & 23.60                     & 57.55                     & 76.82                     & 35.09                     & 37.68                     \\
\midrule
CL4SRec                & 38.66                     & 48.22                     & 26.96          & 28.23          & 37.10                     & 48.93                     & 23.95                     & 25.52                     & 61.08                     & 78.99                     & 38.48                     & 40.88                     \\
DuoRec                 & \underline{40.26}               & \underline{50.13}               & \underline{28.39}    & \underline{29.71}    & \underline{38.87}               & \underline{50.95}               & \underline{25.36}               & \underline{26.96}               & \underline{63.06}               & \underline{82.07}               & \underline{40.24}               & \underline{42.78}               \\
\midrule
FMLP-Rec               & 39.38                     & 48.47                     & 27.85          & 29.06          & 37.23                     & 48.86                     & 24.33                     & 25.87                     & 61.17                     & 80.37                     & 37.97                     & 40.56                     \\
\midrule
Recommender            & 35.14                     & 45.96                     & 22.22          & 23.66          & 33.70                     & 46.40                     & 20.38                     & 22.06                     & 60.33                     & 79.08                     & 36.52                     & 39.03                     \\
\ac{STEAM}             & \textbf{42.09}\rlap{$^*$} & \textbf{52.21}\rlap{$^*$} & \textbf{28.45} & \textbf{29.81} & \textbf{41.72}\rlap{$^*$} & \textbf{54.82}\rlap{$^*$} & \textbf{26.43}\rlap{$^*$} & \textbf{28.17}\rlap{$^*$} & \textbf{66.46}\rlap{$^*$} & \textbf{84.05}\rlap{$^*$} & \textbf{42.83}\rlap{$^*$} & \textbf{45.19}\rlap{$^*$} \\
\bottomrule
\end{tabular}
\end{table*}

\begin{table*}[htbp]
\caption{Robustness analysis of different models. Each value is a performance disturbance.}
\Description{Robustness analysis of different models. Each value is a performance disturbance.}
\label{table_5_7}
\small
\setlength{\tabcolsep}{1mm}
\begin{tabular}{l cccc cccc cccc}
\toprule
                       & \multicolumn{4}{c}{Beauty}                                                & \multicolumn{4}{c}{Sports}                                                & \multicolumn{4}{c}{Yelp}                                                  \\
\cmidrule(r){2-5} \cmidrule(r){6-9} \cmidrule(r){10-13}
Model                  & HR@5             & HR@10            & MRR@5            & MRR@10           & HR@5             & HR@10            & MRR@5            & MRR@10           & HR@5             & HR@10            & MRR@5            & MRR@10           \\
\midrule
GRU4Rec                & -2.21\%          & -1.08\%          & -1.62\%          & -1.35\%          & -2.03\%          & -1.38\%          & -1.96\%          & -1.80\%          & -1.37\%          & -0.91\%          & -1.77\%          & -1.60\%          \\
SASRec                 & -1.67\%          & \underline{-0.68\%}    & -1.81\%          & -1.58\%          & -2.55\%          & -2.10\%          & -2.01\%          & -1.92\%          & \textbf{-0.91\%} & -1.08\%          & \underline{-1.23\%}    & -1.30\%          \\
BERT4Rec               & -2.29\%          & -1.04\%          & -2.52\%          & -2.18\%          & -3.01\%          & -2.96\%          & -4.27\%          & -4.22\%          & -2.81\%          & -2.07\%          & -3.40\%          & -3.19\%          \\
SRGNN                  & -1.85\%          & -1.76\%          & -2.58\%          & -2.53\%          & -1.48\%          & -1.59\%          & -1.96\%          & -1.99\%          & -3.86\%          & -2.71\%          & -4.49\%          & -4.12\%          \\
\midrule
CL4SRec                & -1.60\%          & -1.09\%          & -2.28\%          & -2.12\%          & -2.14\%          & -1.81\%          & -2.36\%          & -2.26\%          & -1.72\%          & -1.46\%          & -2.06\%          & -1.97\%          \\
DuoRec                 & -1.68\%          & -1.28\%          & -1.56\%          & -1.46\%          & -2.34\%          & -1.89\%          & -2.35\%          & -2.25\%          & -1.48\%          & -0.68\%          & -1.49\%          & -1.29\%          \\
\midrule
FMLP-Rec               & \underline{-0.78\%}    & \textbf{-0.51\%} & \underline{-0.57\%}    & \underline{-0.48\%}    & \underline{-1.17\%}    & \underline{-0.93\%}    & \underline{-1.34\%}    & \underline{-1.30\%}    & -1.10\%          & \textbf{-0.48\%} & \textbf{-1.07\%} & \textbf{-0.88\%} \\
\midrule
Recommender            & -1.65\%          & -1.10\%          & -2.71\%          & -2.51\%          & -3.77\%          & -2.89\%          & -4.50\%          & -4.21\%          & -1.76\%          & -1.85\%          & -3.05\%          & -2.96\%          \\
\ac{STEAM}             & \textbf{-0.28\%} & -1.02\%          & \textbf{+0.64\%} & \textbf{+0.44\%} & \textbf{-0.74\%} & \textbf{-0.40\%} & \textbf{-1.20\%} & \textbf{-1.09\%} & \underline{-1.09\%}    & \underline{-0.52\%}    & -1.36\%          & \underline{-1.22\%}    \\   
\bottomrule
\end{tabular}
\end{table*}

\subsection{Overall performance}
\label{section:overall}
To answer RQ1, we compare \ac{STEAM} against the baselines listed in Section \ref{section:baselines} on the real test sets specified in Section~\ref{section:datasets}. 
Table \ref{table_5_1} lists the evaluation results of all methods.
Based on these results, we have the following observations.

First, \ac{STEAM} consistently outperforms all baselines by a large margin on most evaluation metrics of all datasets.
Although \ac{STEAM} only achieves the second best performance in terms of MRR@$5$ and MRR@$10$ on the Beauty dataset, it is almost comparable with the best result.
Compared with the baselines using single recommendation models, \ac{STEAM} jointly trains the recommender and the item-wise corrector, which shares the encoder and item embeddings. 
On the one hand, the deletion and insertion correction tasks for the corrector can provide the recommender with powerful self-supervised signals to obtain better item representations and robust item correlations.
On the other hand, the corrector can correct the input sequence to reduce imperfections so that the recommender in \ac{STEAM} predicts the next item more accurately.
A detailed analysis can be found in the following sections.

Second, the SSL-based models CL4SRec and DuoRec perform better than the vanilla models on all metrics and datasets.
Especially, DuoRec achieves the second best performance on most metrics, and achieves the best performance in terms of MRR@$5$ and MRR@$10$ on the Beauty dataset.
This demonstrates that self-supervised learning can improve the performance of sequential recommendation by deriving extra supervision signals from item sequences themselves.
The performance of DuoRec is obviously better than CL4SRec, which confirms the effectiveness of the model-level augmentation method and the sampling strategy proposed in DuoRec.

Third, although the denoising model FMLP-Rec is an all-MLP model, it shows superior performance compared to the vanilla models that adopt more complex architectures like the transformer.
This is because FMLP-Rec can filter out noisy information in item sequences by \ac{FFT}, while the vanilla models may overfit on the noisy data due to their over-parameterized architectures \cite{lever2016points,mehta2020delight}.
FMLP-Rec performs worse than DuoRec and is comparable to CL4SRec.
Self-supervised learning improves the robustness of DuoRec and CL4SRec to resist the influence of imperfections including noise in item sequences \cite{liu2021contrastive,yu2022self}, so they can fully exploit the power of the transformer to model sequences.

\subsection{Benefits of the corrector}
To answer RQ2, we analyze the effect of the item-wise corrector and the self-supervised correction tasks. 

We first compare the performance of \ac{STEAM} and its recommender in Table \ref{table_5_1}.
We observe that \ac{STEAM} significantly outperforms its recommender in terms of all evaluation metrics on all datasets.
Therefore, we can attribute the improvement of \ac{STEAM} on sequential recommendation to the integration of the recommender with the item-wise corrector.
Moreover, by comparing the recommender of \ac{STEAM} and BERT4Rec, we find that they have almost similar performance.
Because we train the recommender of \ac{STEAM} by the masked item prediction task separately, the recommender of \ac{STEAM} is equivalent to BERT4Rec in this case.

As shown in Table \ref{table_5_2}, we use Overall-R and Overall-C to denote the performances of \ac{STEAM} on the raw item sequences and the corrected item sequences of the real test sets, respectively.
We see that Overall-C is slightly better than Overall-R on most evaluation metrics, but the superiority of Overall-C is not obvious.
We think the reason is that not all test sequences are changed after correction, so we count the percentage of changed test sequences, see Table \ref{table_5_3}.
We can verify that most test item sequences do not change after correction, especially on the Yelp dataset.
Based on the observation in Table \ref{table_5_3}, we group test item sequences into two groups: the changed sequence group whose corrected item sequences are different from the raw item sequences, and the unchanged sequence group where the raw item sequences remain the same after correction.
We evaluate the performance of \ac{STEAM} on the two groups, separately.
As shown in Table \ref{table_5_2}, for the changed item sequence group, we use Changed-R and Changed-C to denote the performance of \ac{STEAM} on the raw item sequences and the corrected sequences, respectively.
We see that Changed-C is better than Changed-R, and the difference between them is bigger than that between Overall-C and Overall-R.
Therefore, we can confirm that the corrector helps the recommender achieve better performance by correcting input sequences.

Next, we compare Overall-R in Table \ref{table_5_2} with all baselines in Table \ref{table_5_1}.
The comparison shows that \ac{STEAM} can significantly outperform all baselines even based on the raw sequences.
We also conclude that the joint learning with the corrector accounts for most of the improvement of \ac{STEAM} over existing methods.
This is because the recommender and the corrector share the encoder and item embeddings.
The item representations and the item correlations learned by the self-supervised deletion correction and insertion correction tasks can be transferred to the recommender, which can enable the recommender to obtain better recommendation results and a robust performance on imperfect sequences.

Finally, we count the percentages of different correction operations performed by \ac{STEAM} during correction.
See Table \ref{table_5_3}.
We observe that \ac{STEAM} chooses to keep items in most cases.
This is reasonable because most item sequences reflect normal user behavior data without noise.
Therefore, \ac{STEAM} tends to keep most sequences and items unmodified.
Based on the statistics, we see that \ac{STEAM} is not only able to delete items but also insert items, which is a key difference from  denoising methods like FMLP-Rec.

We also conduct ablation studies to analyze the effectiveness of each self-supervised task and each correction operation, please see Appendix~\ref{appendix:a} and ~\ref{appendix:b} for details.

\subsection{Robustness analysis}
To answer RQ3, we conduct experiments and analyses on the simulated test sets defined in Section~\ref{section:datasets} by randomly inserting and/or deleting items.
The experimental results are shown in Table \ref{table_5_4}.

The main observations from Table~\ref{table_5_4} are similar to those from Table~\ref{table_5_1}.
The performance for most baselines decreases as most of them cannot handle noisy and missed items.
It is worth noting that \ac{STEAM} achieves better MRR@5 and MRR@10 than DuoRec on the simulated Beauty test set, which indicates that the superiority of \ac{STEAM} becomes more obvious with more imperfect cases.

To further analyze the robustness of different models, we compare their performance on the real test set (see Table~\ref{table_5_1}) and the simulated test set (see Table~\ref{table_5_4}) and calculate the performance disturbance with $dist = (v_{sim}-v_{real})/v_{real}$,
where $dist$ represents the disturbance, $v_{real}$ is the metric value on the real test set, and $v_{sim}$ is the metric value on the simulated test set.
Especially, for \ac{STEAM}, $v_{real}$ is the metric value on the raw item sequences of the real test set (see Overall-R in Table~\ref{table_5_1}), while $v_{sim}$ is the metric value on the corrected sequences of the simulated test set by default.
As we hope to evaluate how \ac{STEAM} handles the simulated imperfections added into the real test set, we consider the performance of \ac{STEAM} on the raw item sequences without correcting the inherent imperfections in the real test set.

The performance disturbance of all models is listed in Table \ref{table_5_7}.
First, we can find that most performance disturbance values are minus, which illustrates that sequence imperfections will degrade model performance.
Second, FMLP-Rec shows competitive robustness and performs better than \ac{STEAM} on the Yelp dataset, which confirms its effectiveness at denoising.
Finally, \ac{STEAM} achieves most of the best results on the Beauty and Sports datasets and most of the second best results on the Yelp dataset, proving its robustness.
The disturbance values of \ac{STEAM} on MRR@5 and MRR@10 on the Beauty dataset are even positive, which illustrates that \ac{STEAM} can not only correct the simulated imperfections but also correct some inherent imperfections.
Although FMLP-Rec is more robust than \ac{STEAM} on the Yelp dataset, it may sacrifice recommendation performance for robustness.
Similarly, SASRec obtains better values on some metrics, but its recommendation performance is relatively inferior.
In contrast, DuoRec is the best baseline in recommendation performance, but its robustness is worse than FMLP-Rec and \ac{STEAM}.
We conclude that \ac{STEAM} strikes a better balance between recommendation performance and robustness than other methods.


\section{Conclusion and Future Work}
We have presented \acs{STEAM}, a \acl{STEAM} that can learn to
correct the raw item sequence before making recommendations by identifying users' misclicks on irrelevant items and/or recalling unexposed relevant items due to inaccurate recommendations.
In order to train the corrector without manual labeling work, we have proposed two self-supervised tasks, the deletion correction and the insertion correction, which randomly insert or delete items and ask the corrector to recover them.
We have conducted extensive experiments on three real-world datasets to show \ac{STEAM} consistently outperforms state-of-the-art sequential recommendation baselines and achieves robust performance on simulated test sets with more imperfect cases.


\ac{STEAM} has the following limitations:
\begin{enumerate*}[label=(\roman*)]
\item it cannot execute `delete' and `insert' on the same item simultaneously; and
\item it can only insert items before the chosen item.
\end{enumerate*}
As to future work, we plan to design a more flexible corrector by repeating the correction process with multiple iterations.
We would also like to combine the corrector with other recommendation models and tasks besides BERT4Rec and the masked item prediction task.

\begin{acks} 
We thank our anonymous reviewers for their helpful comments.
This research was  supported by
the National Key R\&D Program of China with grant (No.2022YFC3303004, No.2020YFB1406704),
the Natural Science Foundation of China (62102234, 62272274, 62202271, 61902219, 61972234, 62072279),  
the Key Scientific and Technological Innovation Program of Shandong Province (2019JZZY010129), 
the Tencent WeChat Rhino-Bird Focused Research Program (JR-WXG-2021411), 
the Fundamental Research Funds of Shandong University, 
and the Hybrid Intelligence Center, a 10-year program funded by the Dutch Ministry of Education, Culture and Science through the Netherlands Organisation for Scientific Research, \url{https://hybrid-intelligence-centre.nl}.
All content represents the opinion of the authors, which is not necessarily shared or endorsed by their respective employers and/or sponsors.
\end{acks}

\bibliographystyle{ACM-Reference-Format}
\bibliography{references}


\clearpage
\appendix
\begin{table*}[ht]
\caption{Ablation study for self-supervised tasks on the real test sets, where \ac{STEAM}-DC is the variant of \ac{STEAM} trained by the deletion correction task and the masked item prediction task, \ac{STEAM}-IC is the variant of \ac{STEAM} trained by the insertion correction task and the masked item prediction task.}
\Description{Ablation study for self-supervised tasks on the real test sets, where \ac{STEAM}-DC is the variant of \ac{STEAM} trained by the deletion correction task and the masked item prediction task, \ac{STEAM}-IC is the variant of \ac{STEAM} trained by the insertion correction task and the masked item prediction task.}
\label{table_7_1}
\small
\setlength{\tabcolsep}{1mm}
\begin{tabular}{l cccc cccc cccc}
\toprule
                       & \multicolumn{4}{c}{Real Beauty} & \multicolumn{4}{c}{Real Sports} & \multicolumn{4}{c}{Real Yelp}            \\
\cmidrule(r){2-5} \cmidrule(r){6-9} \cmidrule(r){10-13}                          
Model                  & HR@5  & HR@10 & MRR@5 & MRR@10  & HR@5  & HR@10 & MRR@5 & MRR@10  & HR@5  & HR@10 & MRR@5 & MRR@10           \\
\midrule
Recommender            & 35.73 & 46.47 & 22.84 & 24.27   & 35.02 & 47.78 & 21.34 & 23.03   & 61.41 & 80.57 & 37.67 & 40.22            \\
\ac{STEAM}-DC          & 41.56 & 51.93 & 27.94 & 29.32   & 41.33 & 54.48 & 26.22 & 27.97   & 66.82 & 83.97 & 43.08 & 45.39            \\
\ac{STEAM}-IC          & 41.77 & 52.23 & 28.15 & 29.54   & 41.19 & 54.06 & 26.14 & 27.85   & 66.87 & 83.75 & 42.96 & 45.36            \\
\ac{STEAM}             & 42.57 & 52.89 & 28.75 & 30.14   & 42.14 & 55.16 & 26.87 & 28.61   & 67.22 & 84.49 & 43.45 & 45.77            \\
\bottomrule
\end{tabular}
\end{table*}

\begin{table*}[ht]
\caption{Ablation study for self-supervised tasks on the simulated test sets.}
\Description{Ablation study for self-supervised tasks on the simulated test sets.}
\label{table_7_2}
\small
\setlength{\tabcolsep}{1mm}
\begin{tabular}{l cccc cccc cccc}
\toprule
                       & \multicolumn{4}{c}{Simulated Beauty} & \multicolumn{4}{c}{Simulated Sports} & \multicolumn{4}{c}{Simulated Yelp}       \\
\cmidrule(r){2-5} \cmidrule(r){6-9} \cmidrule(r){10-13}
Model                  & HR@5  & HR@10 & MRR@5 & MRR@10       & HR@5  & HR@10 & MRR@5 & MRR@10       & HR@5  & HR@10 & MRR@5 & MRR@10           \\
\midrule
Recommender            & 35.14 & 45.96 & 22.22 & 23.66        & 33.70 & 46.40 & 20.38 & 22.06        & 60.33 & 79.08 & 36.52 & 39.03            \\
\ac{STEAM}-DC          & 40.94 & 50.93 & 27.54 & 28.87        & 40.73 & 53.77 & 25.71 & 27.45        & 66.02 & 83.48 & 42.22 & 44.57            \\
\ac{STEAM}-IC          & 40.87 & 51.47 & 27.43 & 28.83        & 40.33 & 53.29 & 25.41 & 27.14        & 65.75 & 83.04 & 41.86 & 44.31            \\
\ac{STEAM}             & 42.09 & 52.21 & 28.45 & 29.81        & 41.72 & 54.82 & 26.43 & 28.17        & 66.46 & 84.05 & 42.83 & 45.19            \\
\bottomrule
\end{tabular}
\end{table*}

\begin{table*}[ht]
\caption{Ablation study for the `delete' operation on the real test sets, where \ac{STEAM}-DK is the variant of \ac{STEAM} that executes `delete' and `keep' operations only.}
\Description{Ablation study for the `delete' operation on the real test sets, where \ac{STEAM}-DK is the variant of \ac{STEAM} that executes `delete' and `keep' operations only.}
\label{table_7_3}
\small
\setlength{\tabcolsep}{1mm}
\begin{tabular}{l cccc cccc cccc}
\toprule
                            & \multicolumn{4}{c}{Real Beauty}      & \multicolumn{4}{c}{Real Sports}      & \multicolumn{4}{c}{Real Yelp}     \\
\cmidrule(r){2-5} \cmidrule(r){6-9} \cmidrule(r){10-13}
\ac{STEAM}-DK               & HR@5   & HR@10  & MRR@5  & MRR@10    & HR@5   & HR@10  & MRR@5  & MRR@10    & HR@5   & HR@10  & MRR@5  & MRR@10 \\
\midrule  
Changed-R                   & 32.08  & 43.56  & 20.01  & 21.53     & 34.89  & 47.46  & 21.48  & 23.15     & 53.36  & 69.32  & 33.35  & 35.63  \\
Changed-C                   & 32.92  & 43.86  & 20.56  & 22.03     & 35.41  & 47.92  & 21.98  & 23.64     & 55.12  & 69.96  & 34.81  & 36.80  \\
\bottomrule
\end{tabular}
\end{table*}

\begin{table*}[ht]
\caption{Ablation study for the `insert' operation on the real test sets, where \ac{STEAM}-IK is the variant of \ac{STEAM} that executes `insert' and `keep' operations only.}
\Description{Ablation study for the `insert' operation on the real test sets, where \ac{STEAM}-IK is the variant of \ac{STEAM} that executes `insert' and `keep' operations only.}
\label{table_7_4}
\small
\setlength{\tabcolsep}{1mm}
\begin{tabular}{l cccc cccc cccc}
\toprule
                            & \multicolumn{4}{c}{Real Beauty}      & \multicolumn{4}{c}{Real Sports}      & \multicolumn{4}{c}{Real Yelp}     \\
\cmidrule(r){2-5} \cmidrule(r){6-9} \cmidrule(r){10-13}
\ac{STEAM}-IK               & HR@5   & HR@10  & MRR@5  & MRR@10    & HR@5   & HR@10  & MRR@5  & MRR@10    & HR@5   & HR@10  & MRR@5  & MRR@10 \\
\midrule  
Changed-R                   & 73.36  & 79.27  & 51.62  & 52.44     & 46.79  & 60.29  & 27.92  & 29.59     & 59.11  & 78.22  & 34.83  & 37.36  \\
Changed-C                   & 75.29  & 80.21  & 55.98  & 56.65     & 47.06  & 61.76  & 28.95  & 31.03     & 60.00  & 79.11  & 35.47  & 38.05  \\
\bottomrule
\end{tabular}
\end{table*}

\section{Ablation study for self-supervised tasks}
\label{appendix:a}
To analyze the effectiveness of each self-supervised task, we carry out experiments with two variants of \ac{STEAM}, i.e., \ac{STEAM}-DC and \ac{STEAM}-IC.
\ac{STEAM}-DC is trained by the deletion correction task and the masked item prediction task.
\ac{STEAM}-IC is trained by the insertion correction task and the masked item prediction task.
The experimental results are shown in Table~\ref{table_7_1} and \ref{table_7_2}.
We observe that both \ac{STEAM}-DC and \ac{STEAM}-IC perform better than Recommender, so we can confirm that using the single deletion correction mechanism or insertion correction mechanism improves the sequential recommendation performance.
We also see that \ac{STEAM} outperforms \ac{STEAM}-DC and \ac{STEAM}-IC, which proves it is necessary to combine these two self-supervised mechanisms to improve model performance.

\section{Ablation study for correction operations}
\label{appendix:b}
To evaluate the effectiveness of each correction operation, we carry out experiments with two variants of \ac{STEAM}, i.e., \ac{STEAM}-DK and \ac{STEAM}-IK.
\ac{STEAM}-DK executes `delete' and `keep' operations, while \ac{STEAM}-IK executes `insert' and `keep' operations.
We focus on the changed sequence group, and report the Changed-R and Changed-C of \ac{STEAM}-DK and \ac{STEAM}-IK on the real test sets.
The results are shown in Table~\ref{table_7_3} and \ref{table_7_4}.
The result of Changed-C are better than those of Changed-R, in both tables.
It illustrates that both deleting items and inserting items are useful, which can make the corrected sequence better for sequential recommendation.
Moreover, the results of Changed-R in Table~\ref{table_7_3} and \ref{table_7_4} are different, which suggests that the sequences corrected by the `delete' operation are different from those corrected by the `insert' operation.
Therefore, \ac{STEAM} should employ both `delete' and `insert' operations to correct item sequences.

\end{document}